# Observing Open Clusters with a Sequence of Ages with *Kepler*

*A White Paper in response to NASA call for community input for alternative science investigations for the Kepler spacecraft*

Submitted 3 September 2013

Joyce A. Guzik[1], Paul A. Bradley[1], Robert Szabo[2], Laszlo Molnar[2], Andrzej Pigulski[3], Konstanze Zwintz[4], Chow-Choong Ngeow[5], Jesper Schou[6], Gerald Handler[7]

1) Los Alamos National Laboratory, Los Alamos, NM USA (joy@lanl.gov)
2) Konkoly Observatory, Budapest, Hungary
3) Astronomical Institute, Wroclaw University, Wroclaw, Poland
4) Instituut voor Sterrenkunde, KU Leuven, Leuven, Belgium
5) Graduate Institute of Astronomy, National Centeral University, Jhongli Taiwan
6) Max Planck Institute for Solar System Research, Katlenburg-Lindau, Germany
7) Nicolaus Copernicus Astronomical Center, Warsaw, Poland

**Abstract:** We propose to observe with *Kepler* an age sequence of nearby uncrowded open clusters. The current *Kepler* field contains very few (only 4) somewhat distant and/or old clusters. Nearby open clusters are already well characterized from ground- and many space-based observations. Our proposal focuses on mid and upper main-sequence variables (main sequence and pre-main sequence gamma Dor or delta Scuti stars, SPB/beta Cep stars, Cepheids, or yellow supergiants), having periods of hours to days accessible by longer-cadence observations. Asteroseismology for these objects to date is limited by the number of modes observable from the ground, difficulty in obtaining spectroscopic or photometric mode ID for the fainter targets that have been observed by *Kepler*, uncertainties in interior and initial abundances, especially for stars with abundance anomalies, uncertainties in distance/luminosity, and lack of knowledge of prior evolution history. The additional constraints of common age, distance and initial abundances in clusters will place these variables in their evolutionary context, and help unlock some of the science and reduce uncertainties for understanding star formation and stellar evolution.

*Introduction and Motivation*

Stellar cluster observations are extremely valuable since clusters contain a stellar population with common age, distance, and initial element abundance that provide additional constraints for stellar models. We propose to maximize the return for stellar data by targeting open clusters with a sequence of ages. The stellar field observed to date with *Kepler* contains only four distant (> 1000 pc) and/or old (e.g. > 1 Gyr) open clusters, NGC 6791, NGC 6811, NGC 6819, and NGC 6866. Younger and/or more nearby clusters will contain variables on the mid- to upper-main sequence that are pulsating in a few to hundreds of simultaneous modes, both radial and nonradial, with easily detectable amplitudes, with periods of hours to days, that are amenable to relatively long cadence integrations and short time series observations.  Nearby open clusters can also have relatively uncrowded fields, making it possible to disentangle the photometric data in cases where targets are moving between pixels during an observing series, or after a spacecraft rotation.

Hundreds of new variables have already been discovered by *Kepler* using long-cadence observations, including gamma Doradus and slowly pulsating B (SPB) g-mode pulsators as well as delta Scuti and beta Cephei p- and mixed mode pulsators.  Somewhat longer period and higher-amplitude Cepheids, RV Tauri, or yellow or blue supergiant pulsators can also be discovered or observed with high precision by *Kepler*.  Long-cadence *Kepler* observations are able to track rotation and starspots on magnetically active stars.  Many eclipsing binaries and contact binaries with periods of hours to days have also been discovered.  The constraints provided by rotation, eclipsing binaries, and multimode pulsations can be combined with the cluster constraints on age, metallicity and distance.  Brighter open clusters have been the subject of extensive ground-based spectroscopic and multicolor photometric observations, providing accurate effective temperatures, surface gravities, element abundances, and rotational and orbital velocity observations for many stars in the clusters.  The nearby stars will be accessible for high-resolution spectroscopy and long time-series multicolor photometry for mode identification that is critical to decipher the asteroseismic content for these stars.

Some stars are known to be variable in many of these clusters; in addition to discovering new variables, it would also be extremely useful to see what we are missing in the known variables.  Are there lower amplitude or longer-period modes that have not been discerned with ground-based photometry? Are there "hybrid" beta Cephei/SPB and delta Scuti/gamma Doradus stars in open clusters and can we use the cluster constraints to place them in evolutionary context? How many stars show photometric signs of starspots or stellar activity?  *Kepler* can help to answer these questions.

In addition to targeting nearby clusters with an age sequence and interesting science objectives, we propose to target clusters that are not too centrally condensed, and do not have a large number of contaminating field stars, so as to avoid overcrowding and facilitate deciphering the stellar content as the stars traverse CCD pixels.  We would like to target clusters accessible to telescopes from both hemispheres to maximize use of ground-based facilities for follow-up and possible future multisite campaigns.

The *Kepler* photometry, even without accurate pointing, will have advantages over the ground-based observations, due to its potential to discover many objects of interest quickly, and its unprecedented signal-to-noise photometry, enabling the detection of modes with several hundred ppm amplitude in just a few days of observing, and the lack of 1 cycle/day alias and gaps that can otherwise be minimized only by a large time-consuming and difficult to organize coordinated campaign of multisite observations.

*Science Justification for Observing Open Clusters*

Some of the science accessible by observing one or more clusters with a range of ages and metallicities includes the following:

1) We can observe pulsating *pre-main sequence* stars with *Kepler* for the first time. No pre-main sequence stars were observed in the present *Kepler* field of view. Many clusters are known to have pre-main sequence gamma Dor and delta Sct stars, and we are just beginning to exploit their potential for asteroseismology and testing pre-main sequence stellar evolution and pulsation modeling.
2) A-F stars have many anomalies; many have peculiar surface abundances with enhancements of some elements and depletion of others; some have low metallicity (lambda Boo stars), some are rapidly rotating, others slowly rotating. Depletion of lithium indicates mixing from rotation and/or convection; Li depletion is an excellent proxy for stellar age, if the depletion mechanisms are well understood and can be calibrated. Mass loss, radiative levitation, diffusive settling, rotational mixing, and convection are competing processes to alter composition stratification and surface abundances. Some A-F stars show no pulsations even though they have effective temperatures and surface gravities placing them in the theoretically expected instability region, while others show gamma Dor or delta Scuti or hybrid behavior but are too cool or too hot for current theory to explain the driving of these modes. The *Kepler* data has shown that, in the F-G star range, there is an ill-defined transition between g-mode and solar-like pulsators, and between pulsating and magnetically active stars; examining stars within the same cluster may help clarify this picture.
3) Additional science questions we could answer include: Why are there many apparently constant stars within instability strips, and what separates them from the pulsators? What limits pulsation amplitudes? Can we verify a proposed new class of variable stars that lies between the SPB and beta Cep instability regions? Can we establish a clearer relationship between metallicity and pulsation properties and instability strip boundaries?
4) Asteroseismology of many stars in a series of cluster ages with different metallicities would provide constraints to help us test and refine stellar model physics at an unprecedented level of precision. We have many questions about stellar opacities, convective overshooting, interior element abundances, gravity-mode mixing, angular momentum transport, magnetic fields, and interior differential rotation that such data would help us to answer. This data will be especially useful to validate the new generation of 2D and 3D stellar evolution and pulsation models that are now becoming feasible.
5) Some clusters with nearly the same ages (e.g. Hyades and Praesepe) show diversity of stellar population, even within the cluster. Intercomparing the stellar content of such clusters will help shed light on star formation processes, e.g. the possibility of episodes of star formation within a cluster.
6) A definitive explanation for blue stragglers has yet to be found. Some blue stragglers are in the delta Scuti instability region and should be pulsating. More constraints and observations may help us confirm their nature and their origin (e.g. binary mergers vs. extra core mixing to extend their lifetimes).
7) SPB/beta Cephei stars: The predicted instability strips of these stars depend sensitively on metallicity and opacities and convective overshooting. Models with a unified parameter set have been unsuccessful to date in predicting the pulsation properties of these stars. Additional constraints provided by clusters should help to resolve this problem.
8) Cepheids: Cepheids lie in the 'blue loop' portion of evolution in H-R diagrams and are core-He burning; these blue loops are not well modeled, and depend sensitively on opacities, abundances, metallicity, and nuclear reaction rates. Improved constraints from clusters would help to constrain model physics.
9) Very few white dwarfs have been found in the *Kepler* field so far. White dwarfs, perhaps even pulsating white dwarfs, might be discovered in open clusters.
10) Observations targeting brighter cluster stars would give us an advance 'preview' and useful experience for observing strategies and target selection in the upcoming TESS mission.

*Proposed Possible 29 Open Clusters and Science Considerations*

The following table summarizes possible target clusters, their age, location, and special features.

| | **Cluster Name** | **Constellation** | Dec (deg) | Ecliptic Latitude (deg) | **Age (My)** | **Distance (pc)** | **Features** | **Priority** |
|---|---|---|---|---|---|---|---|---|
| 1 | NGC 6231 | Scorpius | -42 | -19 | 3.2 | 1243 | Somewhat distant, but very young, known beta Cep stars, young enough for PMS delta Sct stars | 1 |
| 2 | NGC 6611 (M16, Eagle) | Serpens | -14 | +10 | 7 | 1700 | Very young (PMS stars?) Near both ecliptic and galactic plane, somewhat distant, | 1 |

| | | | | | | | |
|---|---|---|---|---|---|---|---|
| | | | | | | dusty? | |
| 3 | M21 | Sagittarius | -22 | +1 | 12 | 1200 | Very young (PMS stars?) Near both ecliptic and galactic plane, somewhat distant, somewhat sparse and compact | 2 |
| 4 | NGC 869 (h Persei in double cluster) | Perseus | +57 | +40 | 13 | 2079 | Well characterized but distant, compact crowded core, off ecliptic plane | 3 |
| 5 | NGC 884 (chi Persei in double cluster) | Perseus | +57 | +40 | 13 | 2940 | Well characterized but distant, compact crowded core, off ecliptic plane | 3 |
| 6 | NGC 4755 (Jewel Box) | Crux | -60 | -48 | 14 | 1976 | Somewhat distant, but young enough for PMS delta Sct stars, and has known beta Cep stars | 1 |
| 7 | IC2602 (Southern Pleiades) | Carina | -64 | -62 | 30 | 147 | Li depletion ages | 1 |
| 8 | IC2391 | Vela | -53 | -66 | 30 | 147 | Li depletion ages | 2 |
| 9 | M36 | Auriga | +36 | +11 | 30 | 1300 | A little off ecliptic and galactic plane, somewhat distant, fewer members | 2 |
| 10 | M18 | Sagittarius | -17 | +6 | 32 | 1300 | Near both ecliptic and galactic plane, somewhat distant, a little sparse | 2 |
| 11 | Alpha Per (Melotte 20) | Perseus | +49 | +30 | 50 | 200 | Maybe too bright for *Kepler*? | 2 |
| 12 | M25 | Sagittarius | -19 | +4 | 92 | 620 | Near both ecliptic and galactic plane, nearby, known Cepheid U Sgr, a little crowded | 3 |

| 13 | Blanco 1 | Sculptor | -30 | -28 | 100 | 253 | Near ecliptic, nearby | 1 |
| --- | --- | --- | --- | --- | --- | --- | --- | --- |
| 14 | M35 | Gemini | +24 | +1 | 100 | 800 | Near both ecliptic and galactic plane | 1 |
| 15 | NGC 2513 | Carina | -61 | -62 | 110 | 409 | Gamma Dor, delta Sct, and chemically peculiar stars, reasonable spread in sky | 2 |
| 16 | Pleiades (M45) | Taurus | +24 | +4 | 125 | 135 | Near ecliptic plane | 1 |
| 17 | NGC 1647 | Taurus | +19 | -3 | 144 | 540 | Near ecliptic and galactic plane, nearby, known Cepheid SZ Tau | 2 |
| 18 | Wild Duck (M11) | Scutum | -6 | +17 | 250 | 1900 | A little off ecliptic, plane, very rich cluster, but maybe too distant and compact | 1 |
| 19 | M39 | Cygnus | +48 | +57 | 280 | 311 | Off ecliptic plane, but looks very nicely spread, near standard Kepler field? | 1 |
| 20 | M38 | Auriga | +36 | +13 | 290 | 1100 | A little off ecliptic plane, near galactic plane, somewhat distant, richer cluster | 2 |
| 21 | M7 | Scorpius | -35 | -11 | 290 | 300 | A little off ecliptic plane, near galactic plane, very crowded field | 3 |
| 22 | NGC 3532 | Carina | -59 | -56 | 316 | 405 | Age and distance are useful, but off ecliptic | 2 |
| 23 | M37 | Auriga | +32 | +9 | 350 | 1400 | Near ecliptic and galactic plane, somewhat distant, richer cluster | 2 |
| 24 | Coma Star Cluster (Melotte 111) | Coma Berenices | +26 | +4 | 450 | 90 | Near ecliptic plane, interesting age, but maybe too | 1 |

|    |                          |          |     |      |      |      | bright for Kepler?                                   |   |
|----|--------------------------|----------|-----|------|------|------|------------------------------------------------------|---|
| 25 | Hyades (Melotte 25)      | Taurus   | +16 | -5.8 | 630  | 46   | Near ecliptic plane                                  | 1 |
| 26 | Praesepe (M44, Beehive)  | Cancer   | +19 | +1   | 630  | 160  | Near ecliptic plane, similar to Hyades in age, comparison | 1 |
| 27 | NGC 752                  | Andromeda| +38 | +24  | 1800 | 400  | Intermediate age to fill age sequence                | 1 |
| 28 | M67                      | Cancer   | +12 | -6   | 3600 | 908  | Over 30 blue stragglers, near ecliptic               | 1 |
| 29 | NGC 188                  | Cepheus  | +85 | +65  | 7000 | 1660 | Oldest open cluster, uncrowded field                 | 2 |

Summary of Properties of 29 Clusters:
Priority 1: 16
Priority 2:  10
Priority 3:  3
<±30 of ecliptic plane 21
< 1000 pc:  17

*Observational Strategy Considerations*

The above table gives a list of possible cluster targets.   21 out of 29 are within 30 degrees of the ecliptic plane.  We understand that a location at or near the ecliptic plane is ideal for maintaining pointing up to one arc second accuracy, and keeping the target on the same pixel for a long time period (weeks to months).  This constraint argues that we should keep as high priority targets near the ecliptic plane.

However, the cluster distance, observing a series of ages, known variable population, and lack of crowding may be more important considerations for the science purposes.  If a cluster is uncrowded, then drifts of a star through many pixels over the course of an observing series may be relatively easily taken into account.  We understand that the spacecraft will need to be rotated for off-ecliptic observations every ~1 day, and the stars will then fall on different pixels.  If the stars sweep through pixels or are rotated to another part of the CCD array each day, the data will need to be normalized/intercalibrated between pixels; we may not be able to study absolute pulsation amplitudes with *Kepler* in this operation mode, but our main objective is to obtain pulsation frequencies via Fourier transforms, and this goal would not be compromised as much by such data.  Moreover, the pulsation periods for many of the stars we are observing are less than 1 day, so we will have continuous series between rotations of one to several pulsation periods.

We would like to observe a given cluster for at least a month to optimize return for longer-period pulsators, binaries, and active stars, but even a week of observations per cluster will yield valuable data.

Masks and Cadence:  We are not proposing to use full-frame images.  Instead, we would like to use a few extended masks optimized for the distribution and brightness of stars in a given cluster, similar to the "superstamp" approach of the original mission.   While short cadence (1-minute) is likely not practical and would limit data collection and return for a large number of cluster stars, there would be some utility to using a cadence less than 30 minutes to obtain more light curve points per star per number of pixels and increase the Nyquist frequency.  Perhaps 10 minutes cadence is a good compromise.

While observing a dozen or more clusters with a range of ages and other properties would optimize the science return, execution of even a part of our program would be worthwhile.

### *References*

The literature on open clusters and pulsating variables in open clusters is very extensive.

To research promising target clusters for this white paper, we made use of on-line tools:

1) The Wikipedia list of open clusters ordered by distance, and links/references therein provided a good starting point:
   http://en.wikipedia.org/wiki/List_of_open_clusters
2) The on-line Messier Catalog:  http://messier.seds.org
3) The WEBDA open cluster database maintained by E. Paunzen and C. Stutz, U. Vienna, originally developed by J-C. Mermilliod, including navigation tools to the literature and data for individual clusters, and summaries of several types of pulsating variables in open clusters:  http://www.univie.ac.at/webda/

There are many papers about variable stars of all of the types discussed here, including *Kepler* discoveries/observations, and we give just a few examples:

*Kepler* A-type stars, gamma Dor/delta Scuti stars, and the problems of hybrids and non-pulsating stars: Balona, L., et al., MNRAS,  15, 3531 (2011); Balona, L. and Dziembowski, W.A., MNRAS 417, 591 (2011); Uytterhoeven, K., et al. A&A 534, 125 (2011); Balona et al., MNRAS 414, 792 (2011).

Pulsating B stars:  Balona, L., et al., MNRAS, 413, 2403 (2011); Daszynska-Daszkiewicz, J., et al. MNRAS 432, 3153 (2013); Walczak, P. et al. MNRAS 432, 822 (2013).

Pre-main sequence variables: Zwintz, K., et al. ASSP 31, 269 (2013); Zwintz, K., et al. A&A, 550, 121 (2013).

Cepheids in open clusters: Anderson, R., et al., MNRAS 434, 2238 (2013).

Cepheid blue loops: Valle, G., et al. A&A 597, 1541 (2009).

Eclipsing binaries: Conroy, K.E., et al., arXiv1306.0521C (2013); Gaulme, P. et al., ApJ 767, 82 (2013).

Blue stragglers: Ahumada, J.A. and Lapasset E., A&A 463, 789 (2007).

Li ages: Cargile, P.A. et al., ApJ Letters, 725, 111 (2010); Anthony-Twarog, B.J. et al., AJ, 138, 1171 (2009).

Hyades/Praesepe comparison: Delorme, P. et al., MNRAS 413, 2218 (2011).

TESS: Ricker, G.R. et al., AAS, 215450064, (2010).

Open clusters in current *Kepler* field: Balona, L. et al., MNRAS 430, 3472 (2013); Sandquist, E., et. A., AJ 762, 58 (2013).

A general reference on pulsating variables and asteroseismology:
C. Aerts, J. Christensen-Dalsgaard, and D. Kurtz, *Asteroseismology* (Springer), 2010.